\documentclass[twocolumn,final,aps,prb,showpacs,superscriptaddress,amsmath,amssymb,amsfonts,floatfix,longbibliography]{revtex4-2}
\pdfoutput=1

\usepackage{graphicx}
\usepackage{multirow}
\usepackage{epsfig}
\usepackage{url}
\usepackage{color}
\usepackage{grffile}

\usepackage[colorlinks,breaklinks,bookmarks=true,citecolor=blue,linkcolor=blue,urlcolor=blue]{hyperref}
\usepackage{color}
\usepackage{tabularx}
\usepackage{sidecap}
\usepackage{soul}
\usepackage{float}
\usepackage{amsmath}

\begin{document}

\title{Doping dependence of linear-in-temperature scattering rate in three-orbital Emery model}

\author{Xun Liu}
\affiliation{School of Physical Science and Technology, Soochow University, Suzhou 215006, China}

\author{Mi Jiang}%\,\orcidlink{0000-0002-9500-202X}}
%\email{jiangmi@suda.edu.cn}
\affiliation{School of Physical Science and Technology, Soochow University, Suzhou 215006, China}
\affiliation{Jiangsu Key Laboratory of Frontier Material Physics and Devices, Soochow University, Suzhou 215006, China}
%\date{\today}

\begin{abstract}
Motivated by the recent experimental demonstration of the doping dependence of the linear-in-temperature resistivity coefficient in cuprates, we numerically investigated the three-orbital Emery model focusing on the slope of the $T$-linear electronic and quasiparticle scattering rates by adopting dynamical cluster quantum Monte Carlo simulations. 
Our exploration discovered that the slope of electronic scattering rate evolves linearly with the electron-doping; while it is inversely proportional to the hole-doping level at intermediate doping regime and then crossovers to the linear-like dependence on further hole doping. These features remarkably match with the experimental findings qualitatively. We further discuss the doping-dependent slope of the quasiparticle scattering rate and also estimate the resistivity coefficient.
Our presented work provides promising insight on the three-orbital Emery model, particularly its doping evolution and connection to the underlying mechanism of the $T$-linear resistivity in cuprates.
\end{abstract}

\maketitle

\section{Introduction}
Non-Fermi liquid (NFL) phenomenology, especially its manifestation in the transport property as the so-called ``strange'' metal phase, is ubiquitous in a wide variety of materials~\cite{Rev1,Rev2,Rev3,Rev4,1,2,3,4,5}. One notable example is the linear-in-$T$ resistivity in quite a wide temperature range in the cuprates~\cite{Rev2,Rev3,Rev4,4,5,np}. This distinction from the conventional paradigm of Landau Fermi liquid theory that predicts  $1/\tau \sim T^2$ scattering rate has attracted much attention in the past decades~\cite{Rev2,Rev3,Rev4}, with the significant theoretical progress by the notion of Planckian dissipation~\cite{Zaanen04,Rev3,Zaanen19} as well as a variety of numerical studies mostly in the framework of two-dimensional Hubbard models~\cite{mousatov2019bad,sentef2011superconducting,zlatic2014universal,riera1994optical,pakhira2015absence,dahm1994self,vranic2020charge,cha2020linear,berg2019monte,PhysRevB.94.245134,nfl_pnas,triangular}. In addition, theoretical analysis~\cite{arulsamy2014origin,Shastry3,phillips2011mottness,kiely2021transport,patel2022many,kao2023unified} and even ultracold atomic experiments~\cite{xu2019bad} have provided much insight on these topics as well.

Compared to the single-orbital Hubbard model, the three-orbital Emery model~\cite{emery} provides a more complete description of the copper oxide plane by explicitly including the two ligand O-2p$_{\sigma}$ orbitals in addition to the Cu $d_{x^2-y^{2}}$ orbital in a unit cell. It is therefore generically believed to capture the physics of cuprate SC more accurately than the single-orbital  model~\cite{3b_dqmc,3b_num,nfl_pnas}. 
The rich physics of the Emery model has been extensively investigated via various many-body methodologies~\cite{jiang_pair_2023,chiciak_magnetic_2020,white_doping_2015,mai_orbital_2021,de_medici_correlation_2009,guerrero_quantum_1998,scalettar_antiferromagnetic_1991,kent_combined_2008,sheshadri_connecting_2023,ponsioen_superconducting_2023,vitali_metalinsulator_2019,jiang_density_2023,zhc20,jiang_single-band_2023}, although it remains a challenge of achieving consensus on many aspects of its physics due to the complexity originating from the multi-orbital nature~\cite{zhc20,3b_dqmc,3b_num}.

Previously, we have employed the dynamic cluster quantum Monte Carlo calculations to systematically investigate the temperature dependence of the electronic and quasiparticle scattering rates in the two dimensional three-orbital Emery model~\cite{maotianzong}.
Specifically, our previous simulations~\cite{maotianzong} revealed the linear-in-$T$ scattering rates for a range of intermediate hole dopings. In addition, the small doping levels support the PG behavior while large dopings show the conventional Fermi liquid characterized by $T^2$ scattering rates. This previous study has hinted some interesting doping dependence of the slope of the linear-in-$T$ scattering rates and furthermore its potential relation to the superconducting $T_c$, which demanded more systematical investigation. 

Remarkably, the most recent electronic transport experiments on the cuprate SC have uncovered the relation between the superconducting $T_c$ and the strange-metal's slope $A$ as $T_c \sim \sqrt{A}$~\cite{jinkui}. Besides, the doping dependence of the linear-in-$T$ slope has been shown to exhibit distinct features in hole- and electron-doped sides~\cite{jinkui1,pnas2013,shekhter2024mottphysicsuniversalplanckian}.

\begin{figure}[b]
\psfig{figure=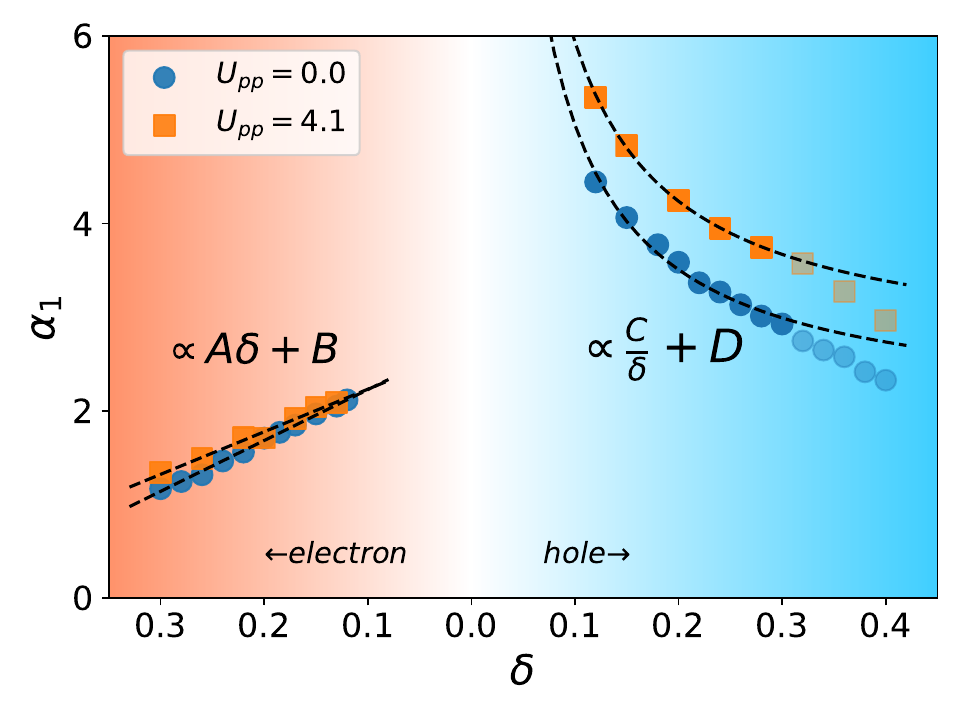,height=6.0cm,width=.45\textwidth, clip}
\caption{Doping dependence of the linear-in-temperature slope $\alpha_1$ of the electronic scattering rate in three-orbital Emery model. The distinct behavior of electron- and hole-doping qualitatively match with the experimental findings~\cite{jinkui,jinkui1}.}
\label{fig1}
\end{figure}

Motivated by these observations, in this work, we conduct detailed exploration on both the electron- and hole-doping dependence of the slope of the scattering rate to provide strong evidence on the asymmetry between two doping sides.
Intriguingly, as illustrated in Fig.~\ref{fig1} as our major result, the linear-in-$T$ slope $\alpha_1$ of the electronic scattering rate has distinct dependence on the electron- and hole-doping. Note that in our hole language $\delta \equiv |\rho-1|$ is the doping level relative to the half-filling $\rho=1$ so that the $\rho<1$ ($\rho>1$) side corresponds to electron- and hole-doping respectively. 
Obviously, the slope $\alpha_1$ shows clear linear evolution on the electron-doping; while it manifests $\sim 1/\delta$ dependence on the hole-doping until around $\delta \sim 0.3$ beyond which the deviated curve switches to a linear-like behavior. The difference between linear and inversely linear dependence is quite remarkable and strongly indicates the asymmetry between electron- and hole-doping. 
Moreover, the additional involvement of the onsite Hubbard interaction $U_{pp}$ of $p$ orbitals do not modify the essential features in the absence of $U_{pp}$.
Therefore, these features are remarkably consistent with the experimental findings qualitatively~\cite{jinkui,jinkui1}.
We thus provide solid numerical evidence on the three-orbital Emery model as the plausible framework of capturing the most important physics of the experimental transport properties.

%%%%%%%%%%%%%%%%%
\section{Model and Method}

The three-orbital Emery model~\cite{emery,3b_dqmc,3b_num} is defined as 
$H=K_0 + K_{pd} +K_{pp}+V_{dd}+V_{pp}$ with
\begin{align}
    K_0 &= (\epsilon_d - \mu)\sum_{i\sigma} n^d_{i\sigma}+(\epsilon_p-\mu)\sum_{j\alpha\sigma} n_{j\alpha\sigma}^p
     \nonumber\\
    K_{pd} &= \sum_{\langle ij \rangle \alpha  \sigma} t_{pd}^{i,j,\alpha}(d_{i,\sigma}^{\dagger}p_{j,\alpha,\sigma}+p^{\dagger}_{j,\alpha,\sigma}d_{i,\sigma})
    \nonumber\\
    K_{pp} &= \sum_{\langle jj' \rangle \alpha \alpha' \sigma} t_{pp}^{j,j',\alpha,\alpha'}(p^{\dagger}_{j,\alpha,\sigma} p_{j',\alpha',\sigma}+p^{\dagger}_{j',\alpha',\sigma} p_{j,\alpha,\sigma})
    \nonumber\\
    V_{dd} &= U_{dd}\sum_{i} n^d_{i,\uparrow}n^d_{i,\downarrow}
    \nonumber\\
    V_{pp} &= U_{pp}\sum_{j,\alpha}n^p_{j,\alpha,\uparrow}n^p_{j,\alpha,\downarrow}
\end{align}
where the hole language is adopted so that $d_{i,\sigma}^{\dagger}$ ($d_{i,\sigma})$ and $p^{\dagger}_{j,\alpha,\sigma}$ ($p_{j,\alpha,\sigma}$) create (annihilates) a hole with spin $\sigma$ in $d_{x^2-y^2}$ orbital at site $i$ and the $p_{\alpha}$ ($\alpha$ = x, y) orbital of site $j$ respectively. $n^d_{i\sigma}$ = $d_{i\sigma}^{\dagger}d_{i\sigma}^{\phantom{\dagger}}$ are the number operators and similar for $n_{j\alpha\sigma}^p$. $\langle . \rangle$ means a sum over nearest-neighbor orbitals. $U_{dd}$ and $U_{pp}$ are the strengths of the $d$ and $p$ on-site interactions, respectively. 
The chemical potential $\mu$ controls the total hole density $\rho$, where $\epsilon_d$ and $\epsilon_p$ are the site energies of the $d$ and $p$ orbitals respectively. $t_{pd}^{ij\alpha}= t_{pd}(-1)^{\eta_{ij}}$ and $t_{pp}^{jj'\alpha\alpha'}= t_{pp}(-1)^{\beta_{jj'}}$ are the nearest-neighbor $d$-$p$ and $p$-$p$ hopping integrals. In the hole language, ${\eta_{ij}}$ and ${\beta_{jj'}}$ take values $\pm$1 following the conventions. In hole language, the phase convention is $\eta_{ij} = 1$ for $j = i + \frac{1}{2}x, \alpha = x$ or $j = i - \frac{1}{2}y, \alpha = y$ and $\eta_{ij} = 0$ for $j = i - \frac{1}{2}x, \alpha = x$ or $j = i + \frac{1}{2}y, \alpha = y$. In addition, $\beta_{jj'} = 1$ for $j' = j - \frac{1}{2}x - \frac{1}{2}y$ or $j' = j + \frac{1}{2}x + \frac{1}{2}y$ and $\beta_{jj'} = 0$ for $j' = j - \frac{1}{2}x + \frac{1}{2}y$ or $j' = j + \frac{1}{2}x - \frac{1}{2}y$, $\alpha = x$ and $\alpha^{'}=y$ or $\alpha = y$ and $\alpha^{'} = x$, respectively~\cite{3b_dqmc}. We adopt the conventional parameters relevant for cuprate SC (in units of eV): $U_{dd}$ = 8.5, $t_{pd}$ = 1.13, $t_{pp}$ = 0.49, $\epsilon_d$ = 0. We will study the effect of finite $U_{pp}=4.1$ compared to the situation of $U_{pp}=0$. 

We employ the dynamical cluster approximation (DCA)~\cite{Hettler98,Maier05,code} with the continuous-time auxilary-field (CT-AUX) quantum Monte Carlo (QMC) cluster solver~\cite{GullCTAUX}.
DCA method maps the bulk lattice problem onto a finite cluster embedded in a mean-field bath in a self-consistent manner~\cite{Hettler98,Maier05}, 
which is accomplished by the convergence between the cluster and coarse-grained (averaged over a patch of the Brillouin zone around a specific cluster momentum $\mathbf{K}$) single-particle Green's functions as well as the self-energies. 
The finite DCA cluster is solved with various numerical techniques, e.g. CT-AUX used here. In principle, increasing the cluster size systematically approaches the exact physical result in the thermodynamic limit. 
Importantly, the finite size of the DCA cluster means that the whole Brillouin zone is approximated by a discrete set of $\mathbf{K}$ points so that the self-energy $\Sigma(\mathbf{K},i\omega_n)$ is a constant function within the patch around a particular $\mathbf{K}$~\cite{Maier05}.

As in our previous work~\cite{maotianzong}, we evaluate the electronic scattering rate $\gamma_{k}\equiv$ -Im$\Sigma^{(2)}(\mathbf{K},\omega = 0)$ by fitting the imaginary part of self-energy -Im$\Sigma(\mathbf{K},i\omega_n)$ at the three lowest Matsubara frequencies, e.g. $n=0,1,2$ to a second order polynomial function of $i\omega_n$ and then extrapolate to zero frequency~\cite{nfl_pnas}.
%Note that, as an approximation avoiding the ambiguous and challenging analytical continuation procedure, the accuracy of this extrapolation for zero frequency improves at low temperatures where the Matsubara frequencies are closer. 
In addition to the electronic scattering rate $\gamma_{k}$, we also investigated the quasiparticle scattering rate or inverse quasiparticle life-time $1/\tau_k = Z_k \gamma_k$, whose difference from $\gamma_{k}$ lies in the incorporation of quasiparticle weight $Z_k$. Both scattering rates are only the approximations of the experimental resistivity. 
%In fact, the quasiparticle picture can even break down in the strange metals~\cite{RevModPhys.94.035004,phillips_stranger_2022,hu_quantum_2022}. Nonetheless, as important physical quantities that can be evaluated efficiently in our numerical DCA technique, they still provide valuable information on the intrinsic physics of the Emery model.

To simulate a wide range of doping levels as well as low enough temperatures with manageable sign problem, we choose DCA cluster $N_c=4$. Although this small cluster loses the information at nodal $\mathbf{K}= (\pi/2,\pi/2)$ direction, we believe that antinodal $\mathbf{K} = (\pi,0)$ direction  encapsulates the most essential physics regarding our interested linear-in-$T$ features of scattering rate.
We note that different $N_c$ can lead to quite similar results for high dopings (large $\rho$) and large $\epsilon_p$~\cite{maotianzong}; while it induces relatively large deviations at low dopings, which has strong momentum dependence of various physical quantities. Fortunately, our focused regime is intermediate to large doping levels to avoid the complicated issues e.g. pseudogap features.

%Another aspect is the momentum differentiation of the scattering rate, namely the deviation between nodal and antinodal directions, which is conventionally associated with the pseudogap (PG) features~\cite{nfl_pnas}. It is also valuable to explore the local scattering rate $\Gamma$ as the momentum average of $\gamma_k$~\cite{triangular}. We believe that this is worthwhile even in the anisotropic situations to explore the difference between momentum averaged scattering rate and the values for a particular $\mathbf{K}$ direction. 

Throughout this work, we examine the slope of the linear-in-temperature evolution of $\gamma_k$ versus the hole density $\rho$ for both electron-doped ($\rho<1$) and hole-doped ($\rho>1$) regimes.
Unlike our previous work exploring two characteristic charge transfer energy $\epsilon_p$~\cite{maotianzong}, here we restrict on the case with $\epsilon_p = 3.24$ eV specific to cuprates.
We mention that, even with the small DCA cluster $N_c=4$, the lowest simulated temperature $T=0.025$ eV corresponds to $\sim 290$ K in reality, which might not be surprising considering the challenge of simulating even simpler single-orbital Hubbard model~\cite{MingpuQin}.

%%%%%%%%%%%%%%%%%%%%%%%%%%%

\section{Results}
In the following sections, we will provide details supporting our major result Fig.~\ref{fig1} and thereby more discussion on the electronic and quasi-particle scattering rates in order. We will finally discuss the approximation of the linear-in-$T$ resistivity by the numerically obtained scattering rates.

\subsection{Electronic scattering rate $\gamma_{k}$}

\begin{figure}
\psfig{figure=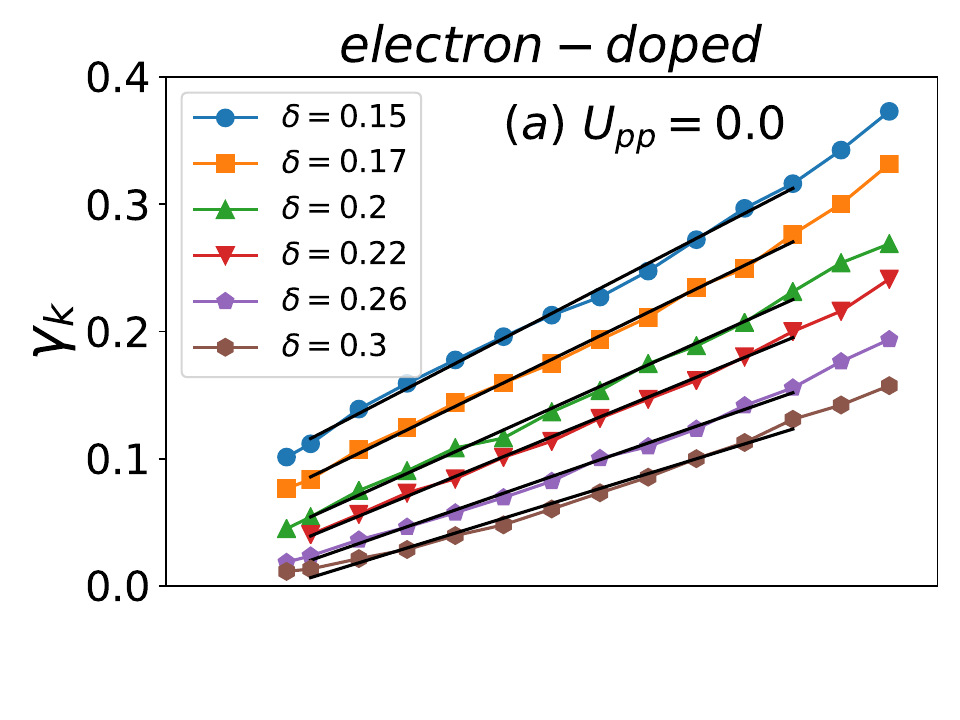,height=3.3cm,width=.24\textwidth,clip}
\vspace{-4mm}
\hspace{-3mm}
\psfig{figure=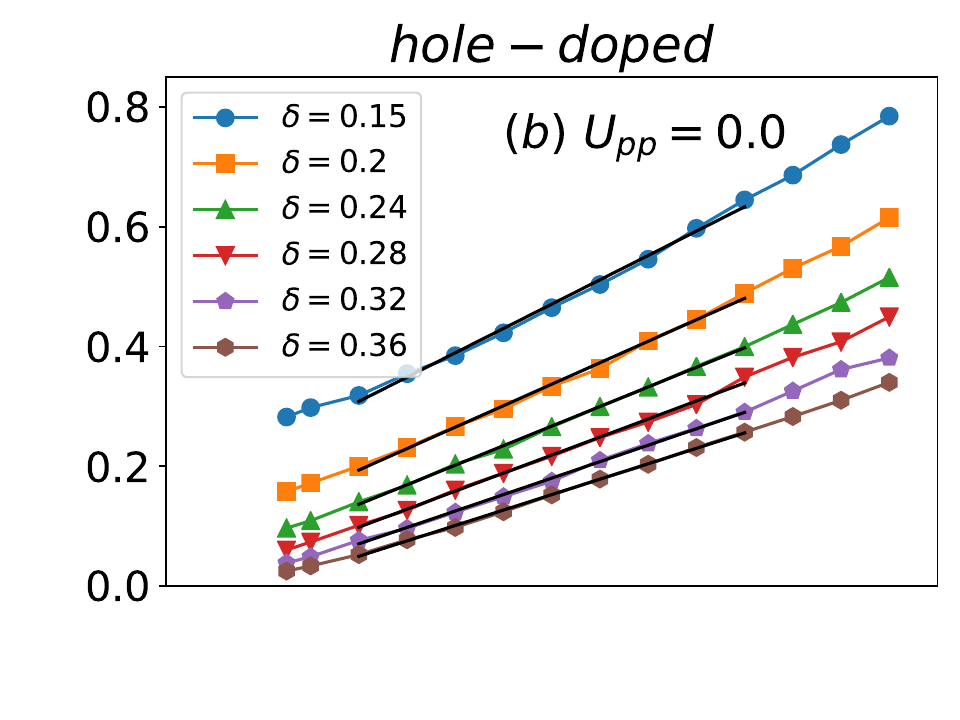,height=3.3cm,width=.24\textwidth,clip}
\psfig{figure=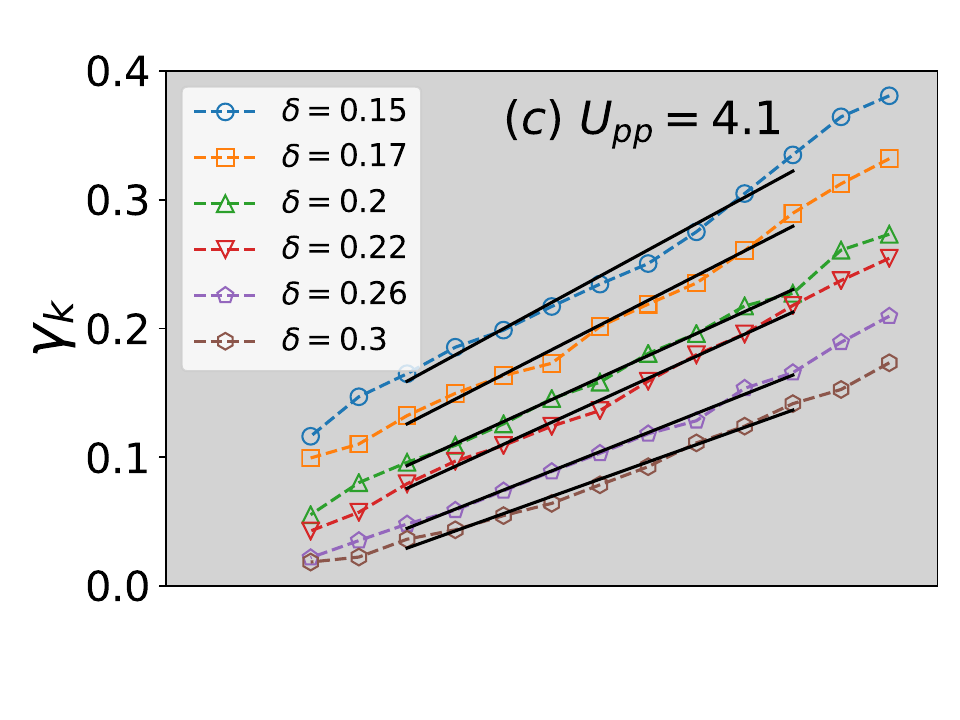,height=3.3cm,width=.24\textwidth,clip}
\vspace{-4mm}
\hspace{-3mm}
\psfig{figure=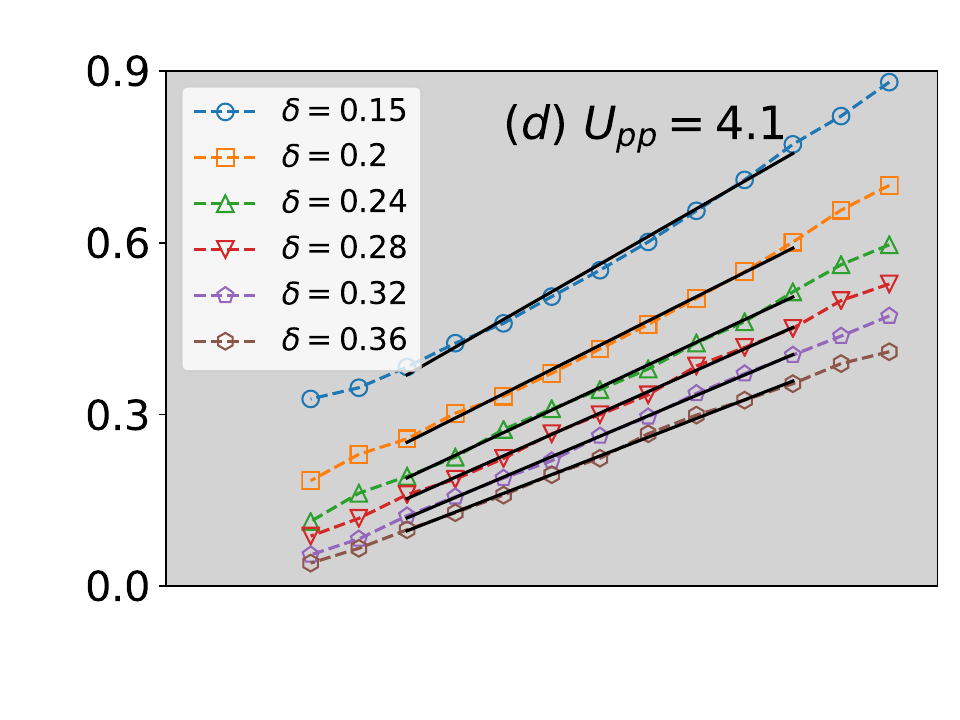,height=3.3cm,width=.24\textwidth,clip}
\psfig{figure=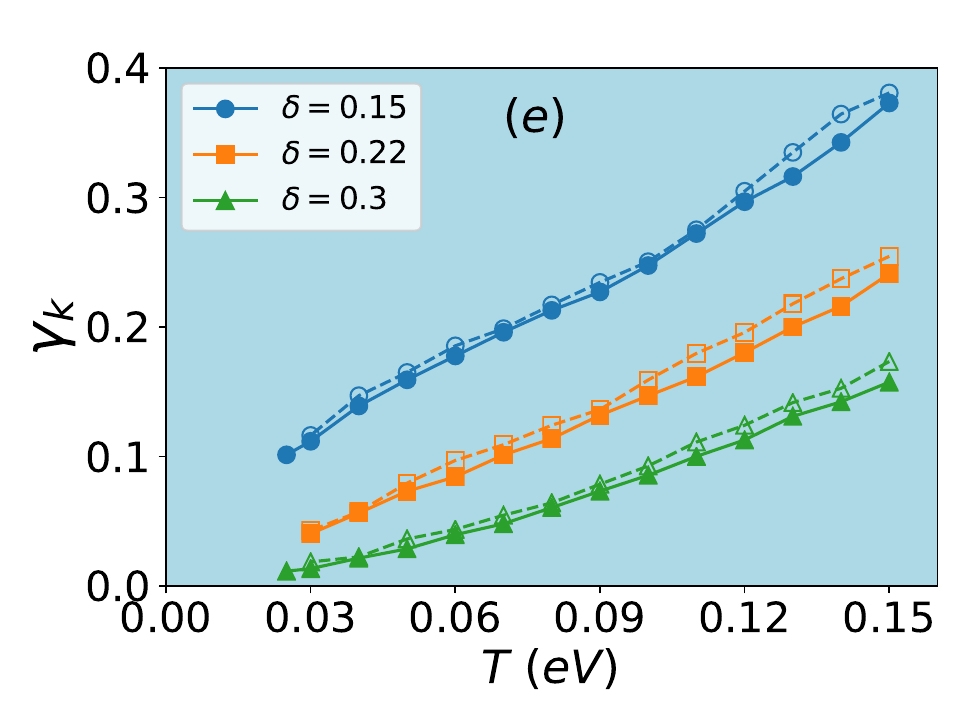,height=3.3cm,width=.24\textwidth,clip}
\hspace{-3mm}
\psfig{figure=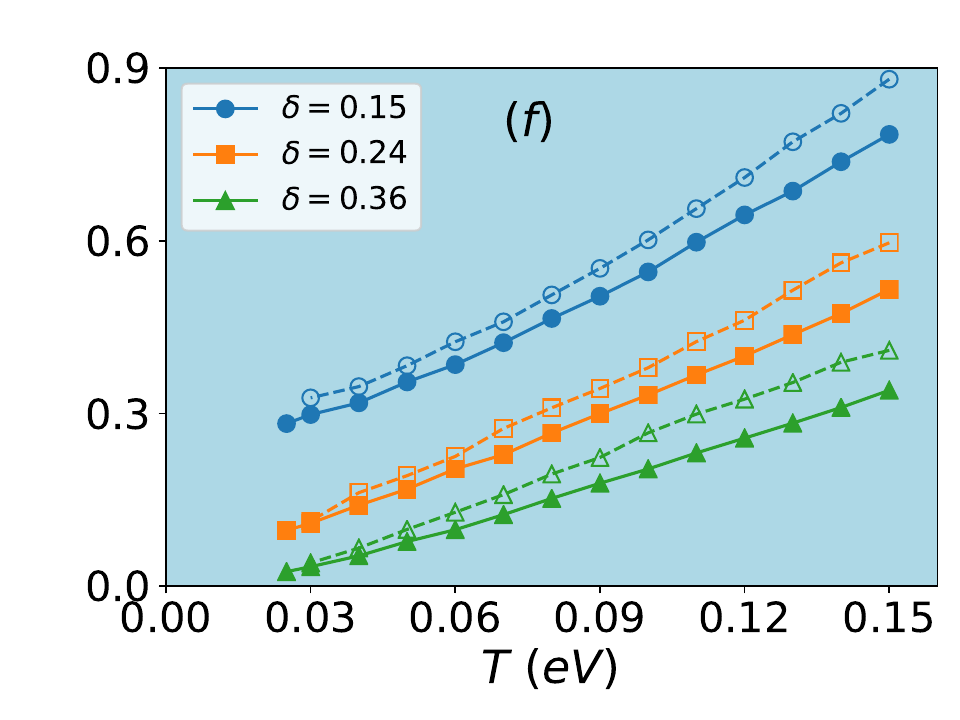,height=3.3cm,width=.24\textwidth,clip}

\caption{(a-d) Temperature dependence of the electronic scattering rate $\gamma_k$ at antinodal $(\pi,0)$ for various electron- and hole-dopings for (a-b) $U_{pp}=0.0$ and (c-d) $U_{pp}=4.1$;
(e-f) Comparison of $U_{pp}=0.0, 4.1$ for selected dopings.}
\label{sigma}
\end{figure}

Fig.~\ref{sigma} illustrates the temperature dependence of antinodal electronic scattering rate $\gamma_k$ of $d$ orbital for various electron- and hole-doping $\delta$. The upper (a-b) and middle (c-d) two panels correspond to $U_{pp}=0.0, 4.1$ separately and the left and right panels are for electron- and hole-doping regimes. 
Here we only list the dopings where the $\gamma_k$ shows clear linear-in-$T$ behavior in a temperature range. In fact, the linear-in-$T$ scattering rate persists for quite a wide density regime at least in our simulated $\rho= 0.7-1.4$, which extends to heavily $\delta=0.3$ electron- and $\delta=0.4$ hole-doped regime.
As discussed in our previous work~\cite{maotianzong}, at small hole doping, for example $\delta=0.05$, the antinodal $\gamma_k$ show a prominent upturn at low temperatures while the nodal direction does not, which is the indication of typical pseudogap (PS) feature. 

As we cannot determine the physical behavior of $\gamma_k(T)$ at even lower temperatures, where the evolution can possibly change in a qualitative manner, therefore, to explore the slope of the scattering rate, we concentrate on the intermediate temperature range, where the $\gamma_k$ shows clear linear temperature dependence. The linear fitting curves are displayed in addition to the original $\gamma_k$ in Fig.~\ref{sigma}. 
Undoubtedly, the slope can depend on the choice of the temperature range for curve fitting, which ultimately affects its doping dependence. We verified that our most important finding in Fig.~\ref{fig1} does not sensitively rely on the fitting procedure so that it is robust against the ``artificially'' chosen temperature range.

As shown in Fig.~\ref{sigma}, with increasing either electron or hole doping, the magnitude of the scattering rate gradually decreases, which indicates the strong metallic nature owing to the heavily doped charge carriers, in spite of its linear-$T$ behavior. 
Besides, the interaction $U_{pp}$ does not affect the generic trend of the evolution of $\gamma_k$ on the doping level.

To have a clear comparison between the two $U_{pp}$ values, Fig.~\ref{sigma}(e-f) display the two curves for some selected dopings. Interestingly, at the electron-doping side, $U_{pp}$ does not have significant impact on $\gamma_k$. In contrast, at the hole-doping regime, adding into $U_{pp}$ apparently enlarges the magnitude of the self-energy indicating the stronger interaction effects. Besides, this strengthen is more obvious at the high temperature regime so that the linear-in-$T$ slope increases accordingly as shown in Fig.~\ref{fig1}.

\subsection{Quasiparticle weight and scattering rate}

Physically, the electronic interaction renormalizes the free electronic properties so that requires the calculation of the quasiparticle scattering rate or inverse quasiparticle life-time $1/\tau_k = Z_k \gamma_k$, where the quasiparticle weight $Z_k$ takes into account the interaction induced dressing of the free charge carriers. Hence, the $Z_k$ value characterizes the closeness to the conventional Fermi liquid (with $Z_k=1$).

\begin{figure}[t]
\psfig{figure=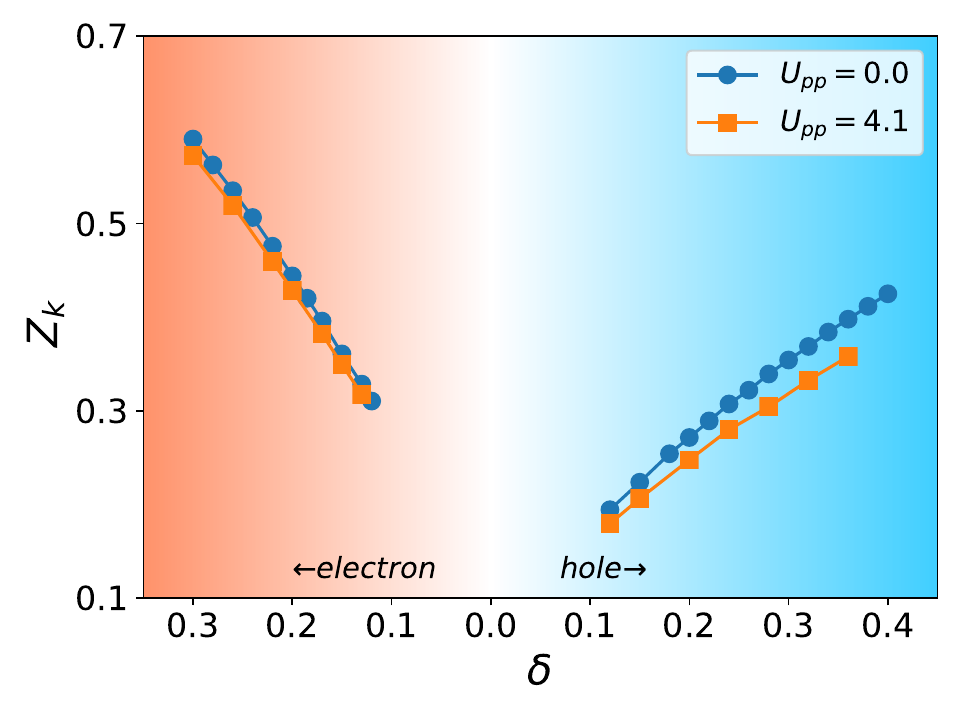,height=6.0cm,width=.45\textwidth, clip}
\caption{Quasiparticle weight $Z(\mathbf{K}=(\pi,0))$ versus the doping $\delta$ for two different $U_{pp} = 0.0, 4.1$ values at the characteristic temperature $T=0.05$.}
\label{Zk}
\end{figure}

In this sense, the quasiparticle scattering rate is indispensable to account for the many-body interaction effects so that is more relevant to the experimental transport properties than the electronic scattering rate $\gamma_k$~\cite{Nolting}. Here the physical picture is based on the assumption that the quasiparticle always survives in our simulations, which should be reasonable for our intermediate to large doping regime. We omit the picture that the quasiparticle can break down in the strange metal phase~\cite{RevModPhys.94.035004,phillips_stranger_2022,hu_quantum_2022}. 

Numerically, $Z_k$ can be approximated by the imaginary part of the self-energy at the lowest Matsubara frequency $\omega_0=\pi T$ as~\cite{Gull2010,triangular}
\begin{align}
    Z(\mathbf{K}) \approx [1-\frac{\Im\Sigma(\mathbf{K},i\omega_0)}{\omega_0}]^{-1}
\end{align}

Because the DCA method is in essence approximating the whole BZ by a discrete set of $\mathbf{K}$ points, here we are only interested in $Z(\mathbf{K}=(\pi,0))$.
Fig.~\ref{Zk} displays that $Z(\mathbf{K})$ increases with either electron- or hole-doping, reflecting the gradual approach to the conventional Fermi liquid. One important feature lies in that, at the same doping level $\delta$, the hole-doping side has generically smaller $Z(\mathbf{K})$ than the electron-doping counterpart. This strongly implies that the hole-doping side has stronger interaction effects. Moreover, the additional consideration of $U_{pp}$ further pushes down the quasiparticle weight but not too much since its impact on $d$ orbital is indirect. However, the effects of $U_{pp}$ is again stronger in the hole-doped regime than the electron-doped side, which is consistent with the Fig.~\ref{sigma}.

\begin{figure}
\psfig{figure=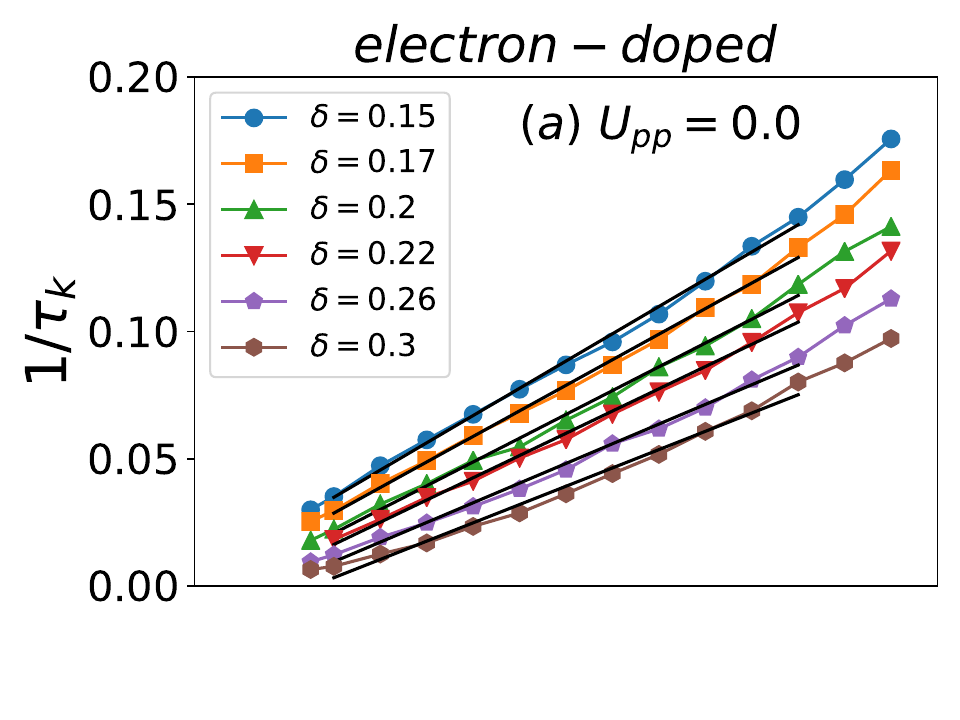,height=3.3cm,width=.24\textwidth, clip}
\vspace{-4mm}
\hspace{-3mm}
\psfig{figure=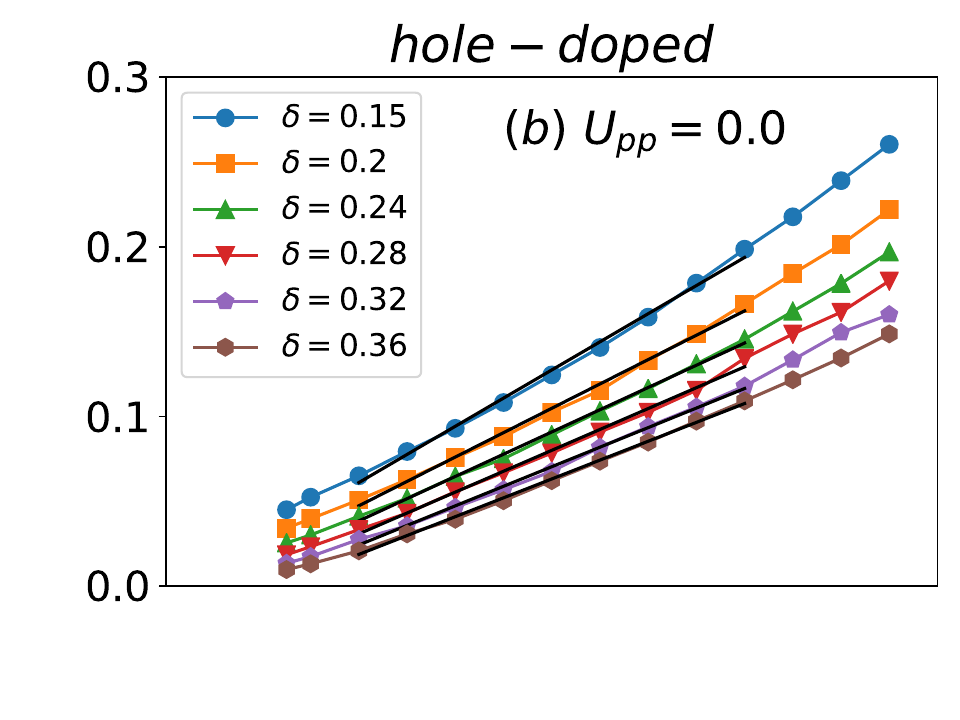,height=3.3cm,width=.24\textwidth, clip}
\psfig{figure=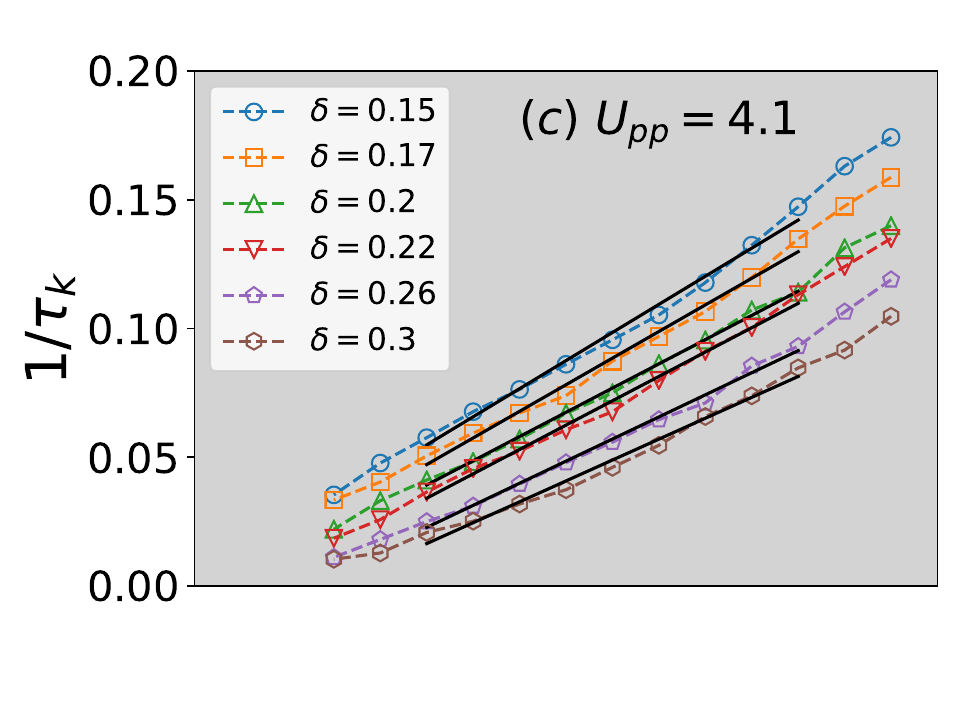,height=3.3cm,width=.24\textwidth, clip}
\vspace{-4mm}
\hspace{-3mm}
\psfig{figure=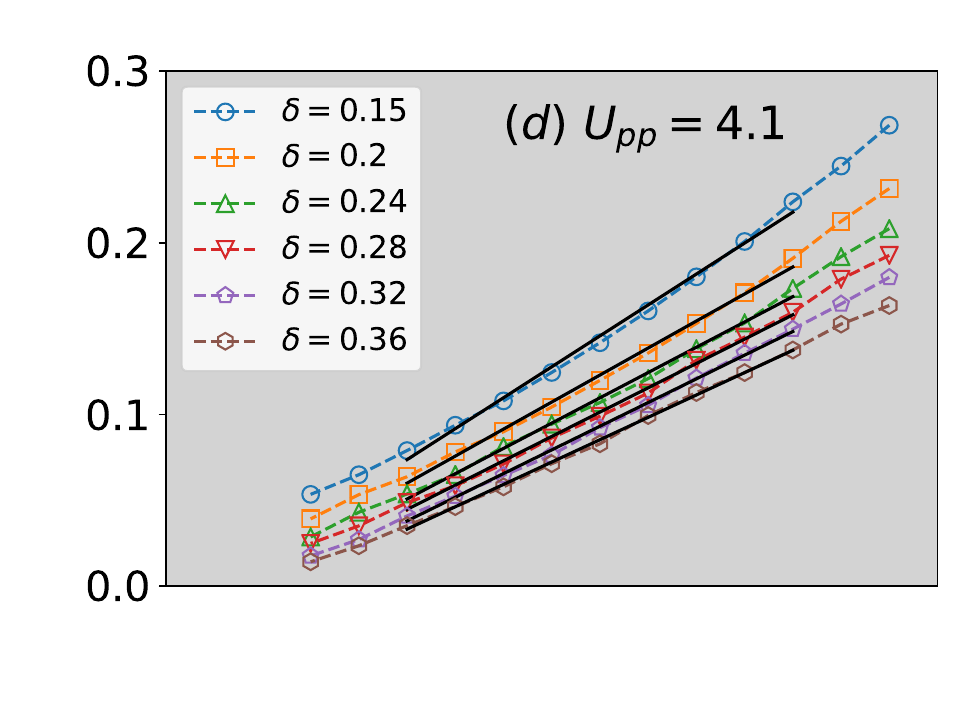,height=3.3cm,width=.24\textwidth, clip}
\psfig{figure=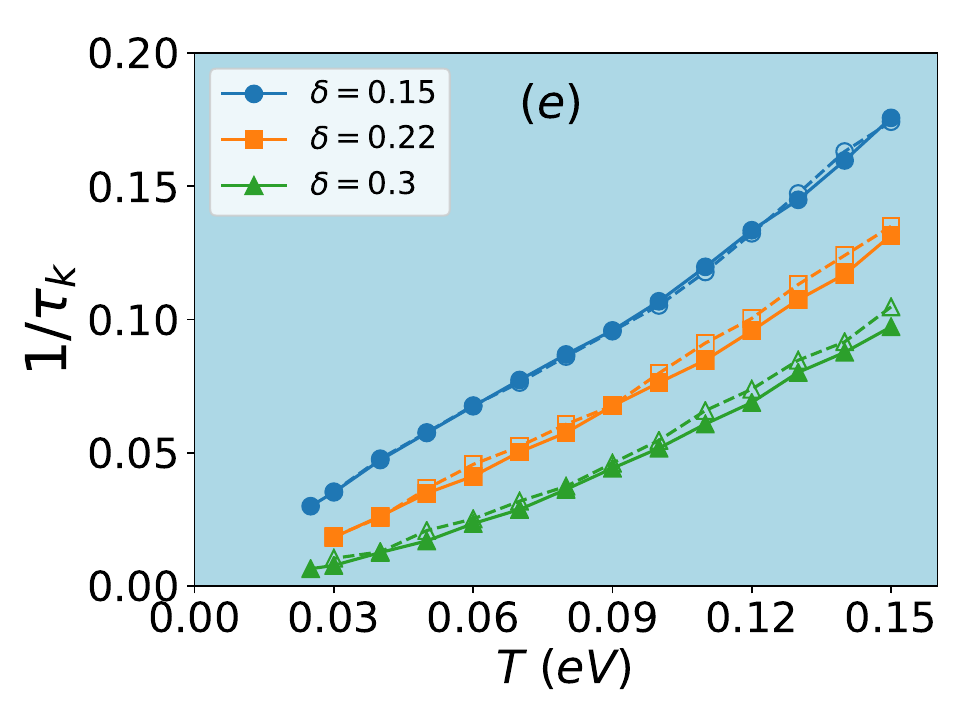,height=3.3cm,width=.24\textwidth, clip}
\hspace{-3mm}
\psfig{figure=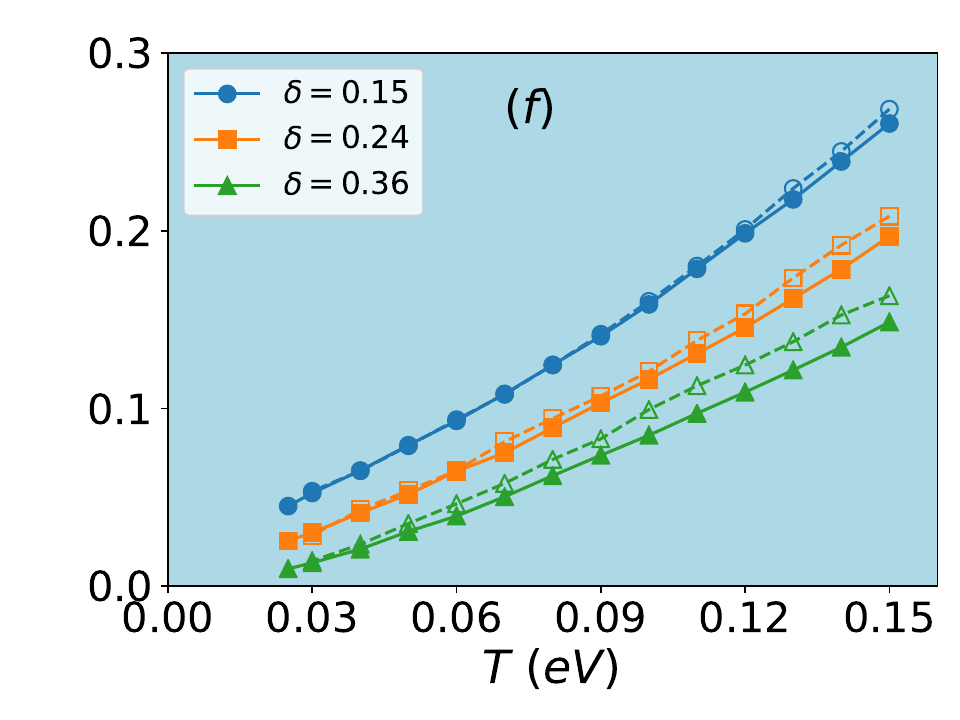,height=3.3cm,width=.24\textwidth, clip}
\caption{The temperature evolution of quasiparticle scattering rate akin to Fig.~\ref{sigma}.}
\label{tauk}
\end{figure}

With the observation of the quasiparticle weight,
Fig.~\ref{tauk} provides the $T$-dependence of the quasiparticle scattering rate, which is qualitatively similar to its electronic counterpart $\gamma_k$.
However, the stronger impact of $U_{pp}$ on the hole-doping side for electronic scattering rate as discussed in Fig.~\ref{sigma} seems weaken for the quasiparticle scattering rate as evidenced by the comparison of panels (e-f).

\begin{figure}
\psfig{figure=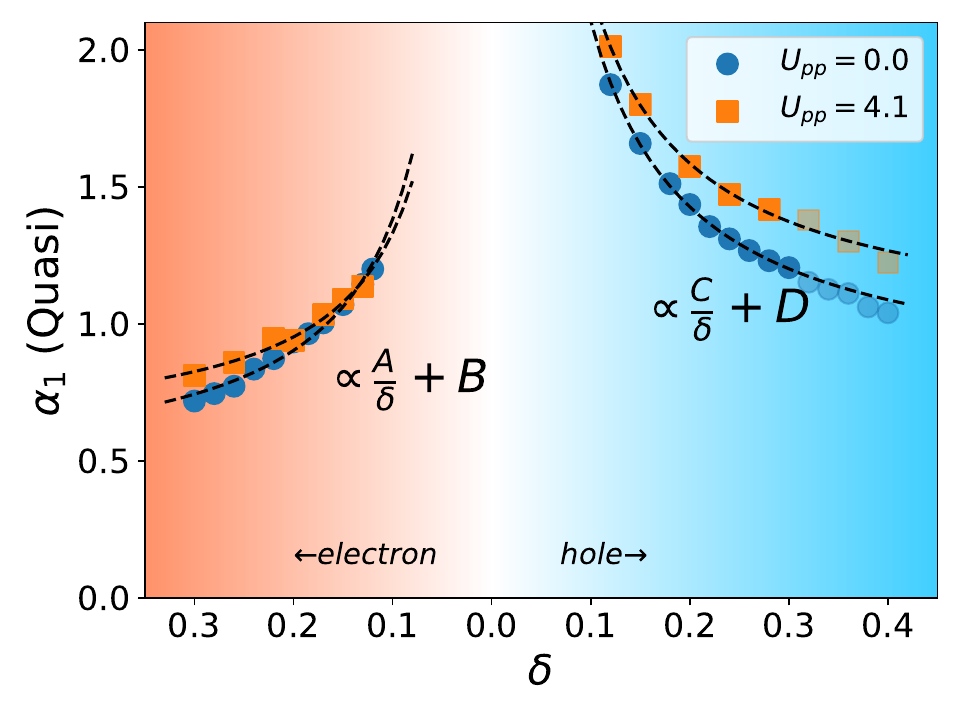,height=6.0cm,width=.45\textwidth, clip}
\caption{Doping evolution of the slope of quasiparticle scattering rate shows the linear dependence in both doping sides with deviation at high $>0.3$ hole-doping, similar to Fig.~\ref{fig1}.}
\label{alphaquasi}
\end{figure}

More importantly, Fig.~\ref{alphaquasi} demonstrates the doping evolution of the slope of quasiparticle scattering rate obtained from Fig.~\ref{tauk}. Obviously, the inversely linear behavior is quite similar to the electronic scattering rate in Fig.~\ref{fig1}. Nonetheless, the electron-doping side switches to an inversely linear feature as well, which is in contrast to the Fig.~\ref{fig1} and also the experiments~\cite{jinkui}.  
This contradiction might be closely related to the numerical approximation of $Z(\mathbf{K})$. In fact, the conclusion can depend on which doping range for fitting. For example, if only taking into account the electron doping $0.1-0.2$ as in experiments~\cite{jinkui}, it can also be roughly fitted as linear doping dependence.
Nonetheless, the generic increase of the quasiparticle $\alpha_1$ by $U_{pp}$ is consistent with Fig.~\ref{fig1}.

Apart from the above features, we have not found any decisive signature for the unity slope associated with the universal Planckian limit~\cite{Zaanen04,Rev3,Zaanen19}, which might indicate the insufficiency of the current approximation of the quasiparticle quantities in the Emery model or even the Planckian theory itself to account for the strange metal phase of the cuprates.

\subsection{Estimation of $T$-linear resistivity coefficient $A_1$}

All the previous discussions are for scattering rates of either electronic or quasiparticle, which are both indirect estimation of the transport resistivity measured in experiments~\cite{jinkui,jinkui1}.
To directly get access to the transport properties, here we follow the assumption and analysis regarding on the relationship $\alpha_1(\delta) \propto \delta A_1(\delta)$ between the $T$-linear slope of scattering rate $\alpha_1$ and the experimental $T$-linear resistivity $A_1$~\cite{jinkui1,Heumen2022}, which allows us to estimate the realistic $A_1$. 

\begin{figure}
\psfig{figure=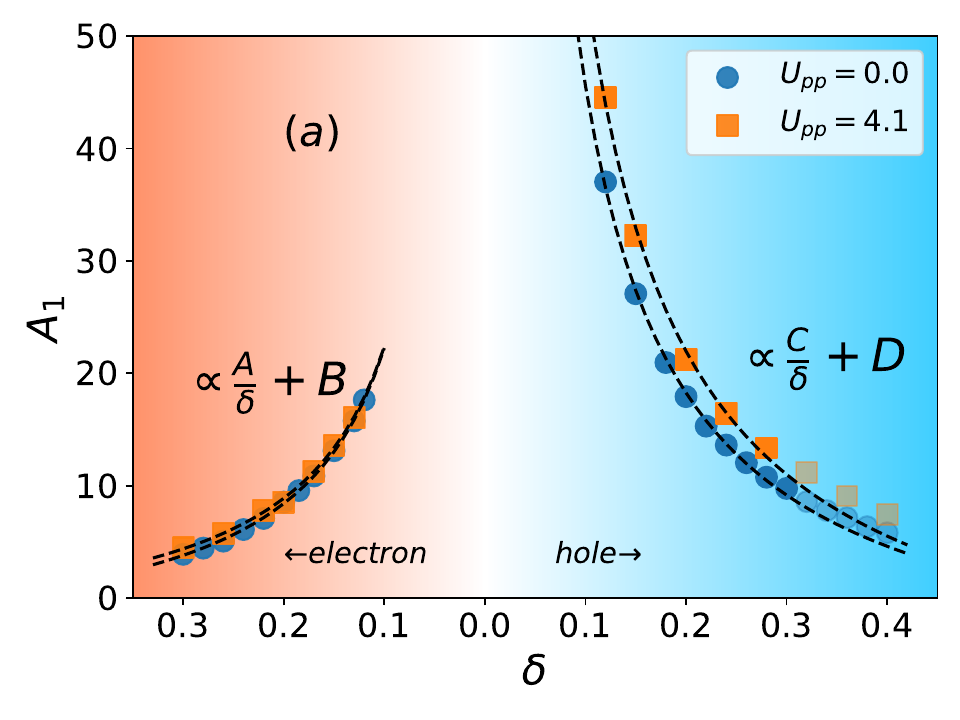,height=6.0cm,width=.45\textwidth, clip}
\psfig{figure=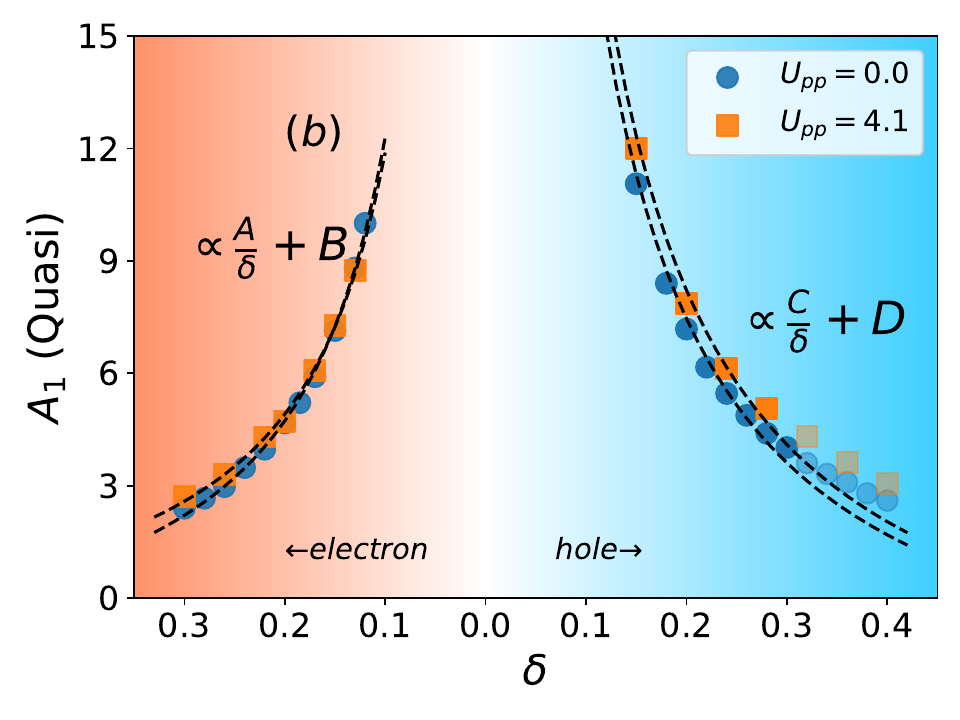,height=6.0cm,width=.45\textwidth, clip}
\caption{Estimation of $T$-linear resistivity coefficient $A_1$ by $\alpha_1(\delta) \propto \delta A_1(\delta)$. The panel (a) is obtained via the electronic $\alpha_1$ in Fig.~\ref{fig1}; while the panel (b) from the quasiparticle $\alpha_1$ in Fig.~\ref{alphaquasi}.}
\label{A1}
\end{figure}

Fig.~\ref{A1} reveals that the estimated $A_1(\delta)$, regardless of panel (a) obtained via electronic $\alpha_1$ in Fig.~\ref{fig1} or panel (b) from the quasiparticle $\alpha_1$ in Fig.~\ref{alphaquasi}, is inversely proportional to both electron- and hole-doping $\delta$. Hence, the hole-doping seemingly agree with the experimental findings while the electron-doping side completely deviates. 

Besides, $U_{pp}$ seems having little impact on the scale of $A_1$ on the electron-doping side as evidenced by the overlapping between two curves. Additionally, the electronic $A_1$ has apparently larger magnitude in the hole-doping side while the quasiparticle $A_1$ manifests more symmetric feature with electron- and hole-doping regime. These features are somewhat obscure, which probably originates from the excessive or oversimplified approximation employed for estimating $A_1$. Whether these features are intrinsic in the Emery model deserves further exploration especially via other advanced many-body techniques.

\subsection{Discussion}

We have illustrated the doping dependence of the linear-in-$T$ slope $\alpha_1$ of the electronic and quasiparticle scattering rate, as well as their corresponding estimated resistivity coefficient $A_1$. Apparently, the $\alpha_1$ of electronic scattering rate in Fig.~\ref{fig1} matches best with the experimental observations. Other quantities either involve subtle approximations or have some intrinsic limitations to be directly compared with the experiments. In this sense, the exploration of the sole electronic scattering rate via numerically evaluated self-energy should be the most straightforward and reliable.

There are more issues deserving elaboration. One issue regards the crossover behavior at around $\rho \sim 1.3$ (hole-doping $\delta \sim 0.3$) in Fig.~\ref{fig1}, which looks disappearing or not obvious in quasiparticle $\alpha_1$ in Fig.~\ref{alphaquasi}
and $A_1$ in Fig.~\ref{A1}. Therefore, it might not be intrinsically connected to any quantum critical doping level. It is unclear whether it is physically realistic, e.g. Lifshitz transition, or simply owing to the numerically fitting procedure so that ``artificial''.

Another issue concerns the relation between the $\alpha_1$ or $A_1$ and more involved superconducting, charge, or magnetic properties, which we have not calculated at present due to the computational cost and inaccuracy associated with these quantities involving two-particle scattering process. 
Physically, the self-energy has included all scattering effects of one added or removed electron originating from the Coulomb interaction among the electrons. Therefore, it is entirely determined by the full two-particle scattering amplitude, which is described by the Dyson- Schwinger equation of motion~\cite{Toschi2015,nfl_pnas}. More sophisticated fluctuation diagnosis along this line should be worthwhile. 

In addition, the direct evaluation of the superconducting $T_c$ should provide more insights.
From the previously estimated $T_c$ showing the roughly quadratic dependence on the doping~\cite{mai_orbital_2021}, it is plausible to expect that the $T_c \sim \sqrt{\alpha_1}$ in the electron-doped regime, although the $\alpha_1$ is only the approximation of the experimental resistivity $A_1$ satisfying $T_c \sim \sqrt{A_1}$ in transport experiments~\cite{jinkui}.

%%%%%%%%%%%%%%%%%%%%%%%%%%%
\section{Summary}

In summary, motivated by the experimental observation of the $T$-linear resistivity coefficient of electron- and hole-doped cuprates~\cite{jinkui,jinkui1}, we have employed the dynamic cluster quantum Monte Carlo calculations to systematically investigate the temperature dependence of the electronic and quasiparticle scattering rates and furthermore the doping dependence of the linear-in-$T$ slope in the framework of two dimensional three-orbital Emery model.

Our numerical simulations reveal that the slope of the linear-in-$T$ electronic scattering rates matches well with the experimental findings. In particular, the slope evolves linearly with the electron-doping; while it is inversely proportional to the hole-doping level at intermediate doping regime and then crossovers to the linear-like dependence on further hole doping. Moreover, we also examined the quasiparticle scattering rate and estimated the linear coefficient of the resistivity, whose doping evolution are somewhat obscure and requires further exploration.

Overall, our numerical findings of the three-orbital Emery model is generally consistent with the transport experiments as expected, which provides evidence on the sufficiency of the Emery model to capture the essential physics associated with the strange metal phase of cuprates.
We anticipate that the presented work provides valuable insight on the Emery model and inspire more work on its properties and more importantly its intrinsic connection with the cuprates. 

\section{Acknowledgement}
We would like to thank Tanusri Saha-Dasgupta and Kui Jin for illuminating suggestion.
We acknowledge the support by National Natural Science Foundation of China (NSFC) Grant No.~12174278 and Priority Academic Program Development (PAPD) of Jiangsu Higher Education Institutions. 

%%%%%%%%%%%%%%%%%%%%%%%%%%%%%
%\bibliographystyle{plain}
\bibliography{main.bib}

\end{document}